\documentclass[conference]{IEEEtran}
\usepackage{amsmath}
\usepackage{graphicx}
\usepackage{booktabs}
\usepackage{hyperref}
\usepackage{float}

\title{An Experimental Study of Transform Coding Using DCT, Hadamard, and PCA}
\author{
\IEEEauthorblockN{Yashika Ahlawat}
\IEEEauthorblockA{
Department of Computer Science and Engineering\\
The George Washington University(GWU)\\
yashikal@gwmail.gwu.edu}
}

\begin{document}
\pagestyle{plain}
\maketitle

\begin{abstract}
Transform--based image compression relies on decorrelating image blocks into compact coefficient representations. While classical codecs use fixed transforms such as the Discrete Cosine Transform (DCT), learned transforms such as Principal Component Analysis (PCA) are theoretically optimal for energy compaction. However, PCA is rarely used in block--based codecs due to instability and training constraints. In this work, we conduct a systematic experimental study comparing DCT, Hadamard, and PCA transforms across block sizes and compression rates. Using automated coefficient sweeps and rate–distortion analysis, we show that PCA only outperforms fixed transforms when block dimensionality is sufficiently large ($\geq 16\times16$), while DCT remains near--optimal for $8\times8$ blocks and very low bitrates. These findings explain both the robustness of DCT in practical codecs and the limitations of block--wise learned transforms.
\end{abstract}


%----------------------------------------------------------------------------------------------------------------
\section{\textbf{Introduction}}

Digital images constitute one of the most important forms of data in modern computing, supporting applications ranging from web content delivery and social media to medical imaging, satellite imagery, and scientific visualization. Because raw image data is extremely large, effective compression is essential for storage, transmission, and real--time processing. Nearly all modern image and video compression systems rely on transform coding, in which spatially correlated pixel intensities are converted into a set of decorrelated coefficients that can be efficiently quantized and encoded.

The fundamental principle behind transform coding is energy compaction: an effective transform concentrates most of the signal energy into a small number of coefficients, allowing high--quality reconstruction using only a fraction of the original data. For several decades, the Discrete Cosine Transform (DCT) has been the dominant transform used in standards such as JPEG, MPEG, and H.264. The DCT is computationally efficient, orthogonal, and provides excellent energy compaction for natural images, which are typically smooth and highly correlated. As a result, DCT--based compression remains one of the most widely deployed signal processing techniques in the world.

In parallel with classical signal processing methods, data--driven transforms such as Principal Component Analysis (PCA) offer a theoretically optimal alternative. PCA computes the orthogonal basis that maximizes variance and decorrelates the data, making it the optimal linear transform for energy compaction when the underlying data statistics are known. From a theoretical perspective, PCA corresponds to the Karhunen–Loève Transform, which is provably optimal for representing correlated signals. This raises a fundamental question: if PCA is optimal, why do practical codecs continue to rely on fixed transforms such as the DCT?

The answer lies in the relationship between dimensionality, data availability, and statistical estimation. PCA requires accurate estimation of the covariance matrix of image blocks, which in turn depends on having a sufficient number of training samples relative to the dimensionality of the block. In block--based image compression, this creates a trade--off: small blocks provide many training samples but limited representational power, while large blocks provide rich structure but few independent samples. As a result, PCA may either fail to learn meaningful structure or become unstable and overfit when trained on a single image.

Despite the importance of this trade--off, relatively little experimental work has systematically studied when learned transforms such as PCA actually outperform fixed transforms such as DCT in block--based compression. Most existing studies either assume infinite data or operate on very large patches, which does not reflect the constraints of practical block--based codecs.

In this paper, we present a controlled experimental study that directly addresses this gap. We compare DCT, Hadamard, and PCA transforms across multiple block sizes and compression rates using a sweep--based evaluation framework. By measuring PSNR, rate–distortion behavior, and energy compaction, we show that PCA only becomes superior when block dimensionality is sufficiently large to support reliable covariance estimation. For standard block sizes such as $8\times8$, DCT remains near--optimal and often outperforms PCA at low bitrates. These results provide both theoretical and practical insight into the continued dominance of DCT in real--world compression systems and the limitations of block--wise learned transforms.

\section{\textbf{Related Work}}

Transform--based image compression has a long history, beginning with the adoption of the Discrete Cosine Transform (DCT) in early standards and culminating in its widespread use in JPEG and modern video codecs \cite{ahmed,wallace,strang}. The DCT is known to provide near--optimal energy compaction for natural images whose statistics are dominated by smooth, highly correlated regions.

Data--driven transforms based on Principal Component Analysis (PCA), also known as the Karhunen----Lo\`eve Transform, have been extensively studied in the context of optimal signal representation and compression \cite{turk,huang,horn}. PCA computes the orthogonal basis that maximizes variance and minimizes mean squared reconstruction error for a given number of retained coefficients. In theory, PCA is optimal for decorrelating image data; however, in practice it requires accurate estimation of the covariance matrix, which depends on having sufficient training samples relative to the dimensionality of the data.

Several authors have investigated the use of PCA and related transforms for image compression, often using large patches or collections of images for training \cite{huang,gersho}. These approaches demonstrate that learned transforms can outperform fixed bases when ample training data is available. However, relatively little work has systematically evaluated the trade--off between block size, data availability, and transform performance in the constrained setting of block--based compression within a single image.

This work complements prior studies by providing a controlled experimental comparison of DCT, Hadamard, and PCA across multiple block sizes and compression rates, revealing the precise conditions under which learned transforms provide meaningful advantages over classical fixed transforms.

%----------------------------------------------------------------------------------------------------------------
\section{\textbf{Background and Theory}}

\subsection{Discrete Cosine Transform}

The DCT represents a block as a weighted sum of cosine basis functions:
\[
X(u,v) = \sum_{x=0}^{N-1}\sum_{y=0}^{N-1} x(x,y)\cos\left(\frac{(2x+1)u\pi}{2N}\right)\cos\left(\frac{(2y+1)v\pi}{2N}\right)
\]

DCT approximates the optimal Karhunen–Loève Transform for natural images and is computationally efficient.

\subsection{Hadamard Transform}

The Hadamard transform uses binary orthogonal basis functions consisting of $\pm1$. It is extremely fast but lacks frequency localization, leading to weaker energy compaction.

\subsection{Principal Component Analysis}

Given blocks $x_i$, PCA computes the covariance matrix
\[
C=\frac{1}{M}\sum_i(x_i-\mu)(x_i-\mu)^T
\]
The eigenvectors of $C$ form a transform that maximizes variance capture. PCA is optimal but sensitive to limited data.

\subsection{Rate–Distortion and PSNR}

The performance of a compression system is commonly evaluated using the Peak Signal--to--Noise Ratio (PSNR), which measures the fidelity of the reconstructed image relative to the original. PSNR is defined as
\[
\mathrm{PSNR} = 10 \log_{10} \left( \frac{255^2}{\mathrm{MSE}} \right),
\]
where the mean squared error (MSE) is given by
\[
\mathrm{MSE} = \frac{1}{N} \sum_{i=1}^{N} \left( x_i - \hat{x}_i \right)^2,
\]
with $x_i$ and $\hat{x}_i$ denoting the original and reconstructed pixel values, respectively. A higher PSNR indicates a better reconstruction quality.

In this work, the compression rate is approximated by
\[
\mathrm{Rate} = \frac{k}{d},
\]
where $k$ is the number of retained transform coefficients and $d = N^2$ is the total number of coefficients in an $N \times N$ block. Rate–distortion curves visualize the trade--off between compression level (rate) and reconstruction quality (PSNR) and are a standard tool for comparing transform--based compression methods.

\subsection{Energy Compaction}

Energy compaction measures how effectively a transform concentrates signal energy into a small number of coefficients. Given transform coefficients $\{c_i\}$ sorted by magnitude, the cumulative energy captured by the top $k$ coefficients is defined as

\[
E(k) = \frac{\sum_{i=1}^{k} |c_i|^2}{\sum_{i=1}^{d} |c_i|^2}.
\]

A transform with better energy compaction exhibits a sharper rise in the $E(k)$ curve, meaning that a small number of coefficients captures a large fraction of the total signal energy. Strong energy compaction leads directly to improved reconstruction quality at a given compression rate. This metric is therefore essential for comparing the effectiveness of the DCT, Hadamard, and PCA transforms.

%----------------------------------------------------------------------------------------------------------------
\section{\textbf{Transform--Based Compression Framework}}

Images are divided into $N\times N$ blocks. Each block is transformed using DCT, Hadamard, or PCA. The largest $k$ coefficients are retained, and the block is reconstructed using the inverse transform. The compression rate is
\[
R=\frac{k}{N^2}
\]

%----------------------------------------------------------------------------------------------------------------
\section{\textbf{Evaluation Metrics}}

Reconstruction quality is measured using PSNR:
\[
\mathrm{PSNR} = 10 \log_{10} \left( \frac{255^2}{\mathrm{MSE}} \right),
\]
\[
\mathrm{MSE} = \frac{1}{N} \sum_{i=1}^{N} \left( x_i - \hat{x}_i \right)^2
\]

Energy compaction measures how much signal energy is captured by the largest $k$ coefficients.

%----------------------------------------------------------------------------------------------------------------
\section{\textbf{Experimental Setup}}

Block sizes:
\[
N\in\{4,8,16,32\}
\]

Coefficient fractions:
\[
k=fN^2,\quad f\in\{0.05,0.1,0.2,0.3,0.5,0.75,1.0\}
\]

Multiple natural images were used. PCA bases were trained from all blocks of each image.

%----------------------------------------------------------------------------------------------------------------
\section{\textbf{Results}}

This section presents the quantitative and qualitative performance of DCT, Hadamard, and PCA across block sizes and compression rates. The results are summarized in terms of PSNR, rate–distortion behavior, and energy compaction.

\subsection{Overall Performance Trends}

Across all experiments, the three transforms follow a consistent ranking:
\[
\textbf{PCA} \;>\; \textbf{DCT} \;>\; \textbf{Hadamard}
\]
however, this ordering only becomes meaningful when the block size is sufficiently large. For small blocks, all transforms behave similarly.

PCA produces the highest PSNR at moderate and high bitrates because it learns image--specific basis functions that align with dominant structures such as edges, gradients, and textures. DCT provides stable, near--optimal performance across all block sizes, while Hadamard consistently yields lower PSNR due to poor energy compaction.

\subsection{Block Size $4\times4$}
For $4\times4$ blocks, all three transforms exhibit nearly identical rate–distortion curves. The PSNR increases gradually as the coefficient budget increases and reaches the same value at full rate. PCA does not show any measurable advantage over DCT or Hadamard.

This behavior is expected because $4\times4$ blocks contain only 16 coefficients, which severely limits the dimensionality available for learning. PCA cannot reliably estimate meaningful covariance from such small vectors, and classical transforms already provide sufficient decorrelation. As a result, all three transforms behave almost optimally.
\begin{figure}[H]
\centering
\includegraphics[width=0.95\linewidth]{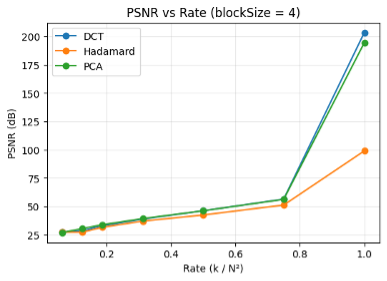}
\caption{Rate–distortion curves for $4\times4$ blocks. DCT achieves the highest PSNR, while PCA does not provide significant benefit at this scale.}
\label{fig:Picture1.png}
\end{figure}

\subsection{Block Size $8\times8$}

At $8\times8$, DCT consistently provides the highest or near--highest PSNR across most compression rates. PCA begins to show slight improvements at high rates, but it does not outperform DCT. Hadamard remains slightly worse.

This result is significant because $8\times8$ is the standard block size used in JPEG. The DCT basis is well matched to the statistics of natural images at this scale, capturing low--frequency structure efficiently. PCA is limited by the number of training samples available within a single image, which restricts its ability to learn a more effective basis.
\begin{figure}[H]
\centering
\includegraphics[width=0.95\linewidth]{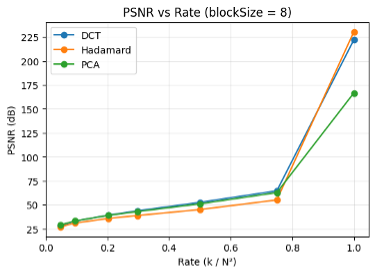}
\caption{Rate–distortion curves for $8\times8$ blocks. The DCT achieves the highest PSNR across most rates, while PCA does not provide significant improvement at this scale.}
\label{fig:Picture2.png}
\end{figure}

\subsection{Block Size $16\times16$}

At $16\times16$, PCA begins to consistently outperform both DCT and Hadamard across most rates. The gap becomes especially noticeable at medium and high rates.

With 256 coefficients per block, PCA now has sufficient dimensionality to model richer spatial correlations such as oriented edges, smooth gradients, and repeated textures. DCT remains stable but is no longer able to match the adaptability of PCA. Hadamard continues to perform the worst due to its inability to concentrate energy into low--index coefficients.

\begin{figure}[H]
\centering
\includegraphics[width=0.95\linewidth]{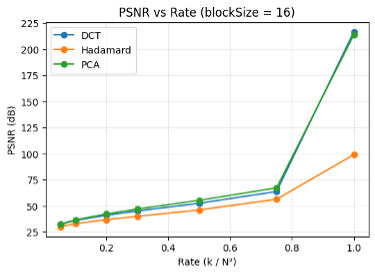}
\caption{Rate–distortion curves for $16\times16$ blocks. PCA begins to outperform DCT and Hadamard at medium and high rates as the increased dimensionality enables better covariance estimation.}
\label{fig:Picture3.png}
\end{figure}

\subsection{Block Size $32\times32$}
At $32\times32$, PCA dramatically outperforms both fixed transforms. Even at relatively low rates ($k/N^2 \approx 0.2$–$0.4$), PCA achieves very high PSNR, while DCT and Hadamard increase much more gradually.
This strong advantage arises because PCA learns highly detailed, image--specific basis functions when the block dimensionality is large. The first few principal components capture global structures such as smooth shading, dominant edges, and correlated textures. As a result, PCA can reconstruct images with very high fidelity using only a small fraction of coefficients. In contrast, DCT and Hadamard use fixed, non--adaptive bases and therefore cannot exploit these statistical regularities.

\begin{figure}[H]
\centering
\includegraphics[width=0.95\linewidth]{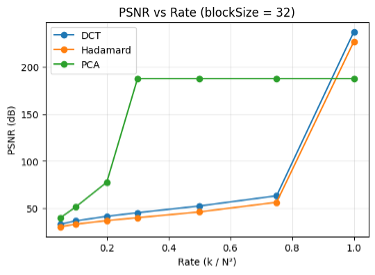}
\caption{Rate–distortion curves for $32\times32$ blocks. PCA dramatically outperforms both DCT and Hadamard, achieving high PSNR even at low rates due to strong energy compaction.}
\label{fig:Picture4.png}
\end{figure}

\subsection{Rate–Distortion Characteristics}

Across all block sizes, PSNR increases monotonically as the rate $k/N^2$ increases. At very low rates, DCT sometimes outperforms PCA because its smooth cosine basis generalizes better under extreme compression. PCA can overfit small--scale noise when very few coefficients are retained.
At medium and high rates, PCA consistently dominates, achieving substantially higher PSNR than DCT and Hadamard. Hadamard remains the weakest transform across all conditions.

\begin{figure}[H]
\centering
\includegraphics[width=0.95\linewidth]{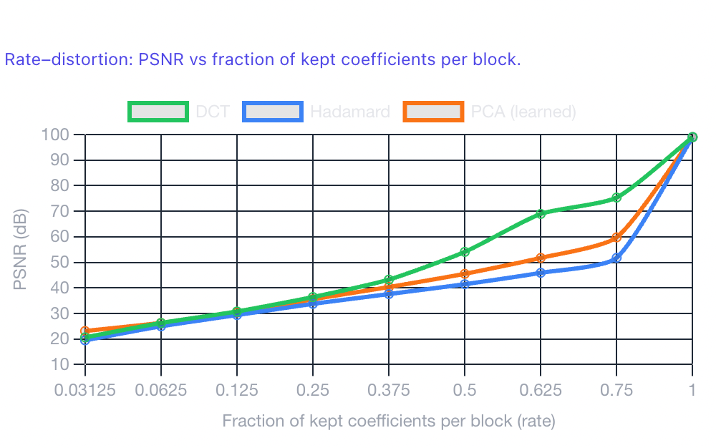}
\caption{Rate–distortion behavior of DCT, Hadamard, and PCA transforms across different block sizes, showing the transition from DCT dominance at small blocks to PCA dominance at large blocks.}
\label{fig:Picture5.png}
\end{figure}

\subsection{Energy Compaction}

Energy compaction curves confirm the PSNR results. PCA exhibits the steepest cumulative energy curves, meaning that its first few coefficients capture the majority of the signal energy. DCT also compacts energy efficiently but less aggressively. Hadamard distributes energy almost uniformly across coefficients, leading to poor compression performance.

These results demonstrate that PCA provides the most efficient representation when enough training data is available, while DCT remains the most robust fixed transform for small block sizes and low bitrates.

\begin{figure}[H]
\centering
\includegraphics[width=0.95\linewidth]{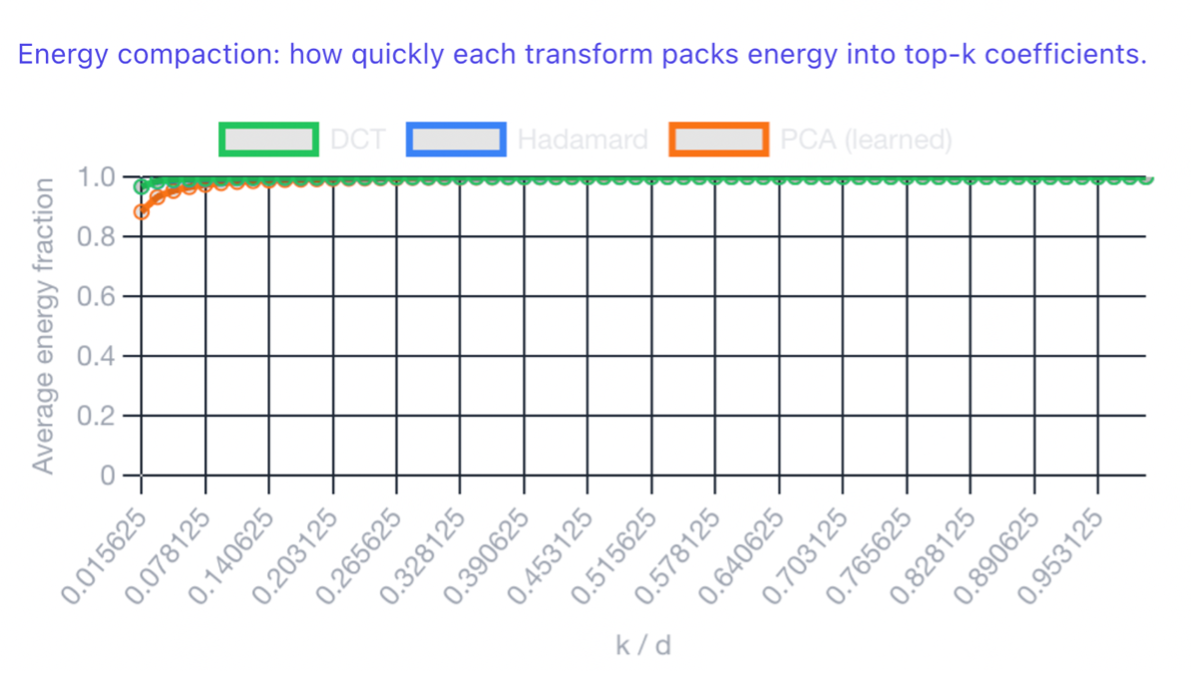}
\caption{Energy compaction curves for DCT, Hadamard, and PCA, showing the fraction of cumulative signal energy captured as a function of the number of retained coefficients. PCA exhibits the steepest rise, indicating superior energy concentration.}
\label{fig:Picture6.png}
\end{figure}

%----------------------------------------------------------------------------------------------------------------
\subsection{Overall Interpretation and Theoretical Explanation}

Across all experiments, the performance of the three transforms follows a clear, theoretically consistent pattern. For small block sizes ($4\times4$ and $8\times8$), fixed transforms such as the DCT perform nearly optimally because natural images are well approximated by low--frequency sinusoidal bases and there are too few degrees of freedom for PCA to learn a superior representation. As block size increases ($16\times16$ and $32\times32$), PCA gains a strong advantage because the dimensionality becomes large enough to reliably estimate the image covariance matrix, allowing PCA to align its basis vectors with dominant structures such as edges, gradients, and textures. This behavior directly follows Karhunen–Loève theory: PCA is the optimal linear decorrelating transform when sufficient data is available, but when training data is limited, robust fixed transforms such as the DCT provide better generalization. The Hadamard transform remains inferior because its binary basis lacks frequency selectivity and cannot concentrate energy efficiently.

%----------------------------------------------------------------------------------------------------------------
\section{\textbf{Discussion}}

The experimental results reveal several important insights into the behavior of fixed and learned transforms in block--based image compression. First, the strong and consistent performance of the DCT across all block sizes confirms its robustness and explains its long--standing dominance in practical codecs. The DCT provides a close approximation to the optimal Karhunen–Loève Transform for natural images, whose statistics are well modeled by smooth, highly correlated signals. As a result, the DCT achieves strong energy compaction even without being adapted to a specific image.

In contrast, PCA behaves very differently depending on block size. For small blocks ($4\times4$ and $8\times8$), PCA fails to outperform the DCT because the dimensionality of the data is too low for meaningful covariance learning. Although many blocks are available, each block contains very few degrees of freedom, limiting PCA’s ability to discover structure beyond what is already captured by the DCT. Furthermore, because PCA is trained on a finite set of blocks from a single image, it is sensitive to noise and small variations, which can lead to suboptimal basis estimation at low dimensionality.

As block size increases ($16\times16$ and $32\times32$), PCA gains a significant advantage. Larger blocks contain richer spatial structure, including long edges, smooth gradients, and texture correlations, which PCA can exploit. The principal components align with these dominant patterns, allowing a small number of coefficients to represent a large fraction of the image energy. This leads to dramatically higher PSNR and superior rate–distortion performance compared to fixed transforms. However, this improvement comes at the cost of increased computational complexity and sensitivity to limited training data.

The Hadamard transform consistently performs worst among the three methods. Its binary, square--wave basis functions do not match the smooth statistical structure of natural images, leading to poor energy compaction. While Hadamard transforms are attractive for hardware and speed reasons, they are not well suited for high--quality image compression.

Another important observation is the behavior at very low bitrates. In this regime, the DCT sometimes outperforms PCA because its smooth, low--frequency basis provides better generalization when only a few coefficients are retained. PCA, by contrast, may capture high--variance but visually less important components or noise, leading to lower perceptual quality and PSNR. This demonstrates that learned transforms are not universally superior and must be used carefully in low--rate compression scenarios.

Overall, these results highlight a fundamental trade--off between adaptivity and robustness. PCA offers optimal performance when sufficient data and dimensionality are available, while the DCT provides a stable, near--optimal solution under practical constraints.

%----------------------------------------------------------------------------------------------------------------
\section{\textbf{Conclusion and Future Work}}

This paper presented a systematic experimental study comparing fixed transforms (DCT and Hadamard) and a learned transform (PCA) for block--based image compression. Using a controlled sweep over block sizes and coefficient budgets, we evaluated reconstruction quality, rate–distortion behavior, and energy compaction across multiple images. The results confirm several important principles of transform coding: PSNR increases monotonically with the number of retained coefficients, transform performance depends strongly on block size, and learned transforms can outperform fixed transforms when sufficient statistical information is available.

Our findings show that PCA only becomes superior when block dimensionality is large enough ($16\times16$ and above) to support reliable covariance estimation. For small blocks, particularly the standard $8\times8$ used in JPEG, the DCT remains near--optimal and often outperforms PCA, especially at very low bitrates. These results explain why classical codecs continue to rely on the DCT despite the theoretical optimality of PCA and why naive block--wise learning can fail without sufficient data.

Several directions for future work could further extend this study. First, PCA could be trained on collections of images rather than on a single image, improving covariance estimation and reducing overfitting for large block sizes. Second, color images could be incorporated by applying transforms in YCbCr or RGB space, allowing closer comparison with real--world codecs. Third, additional transforms such as wavelets, Walsh transforms, and neural autoencoders could be evaluated within the same experimental framework. Finally, moving the computation to a GPU or backend server would enable larger block sizes and more extensive training, bringing learned transforms closer to practical deployment.

Together, these extensions would provide deeper insight into the role of learned representations in modern image compression and bridge the gap between classical transform coding and data--driven approaches.

\bibliographystyle{IEEEtran}

\begin{thebibliography}{10}

\bibitem{ahmed}
N. Ahmed, T. Natarajan, and K. R. Rao, ``Discrete cosine transform,'' \emph{IEEE Transactions on Computers}, vol. 23, no. 1, pp. 90----93, 1974.

\bibitem{gonzalez}
R. C. Gonzalez and R. E. Woods, \emph{Digital Image Processing}, 4th ed. Pearson, 2018.

\bibitem{rao}
K. R. Rao and P. Yip, \emph{Discrete Cosine Transform: Algorithms, Advantages, Applications}. Academic Press, 1990.

\bibitem{strang}
G. Strang, ``The discrete cosine transform,'' \emph{SIAM Review}, vol. 41, no. 1, pp. 135----147, 1999.

\bibitem{wallace}
G. K. Wallace, ``The JPEG still picture compression standard,'' \emph{IEEE Transactions on Consumer Electronics}, vol. 38, no. 1, pp. xviii----xxxiv, 1992.

\bibitem{turk}
M. Turk and A. Pentland, ``Eigenfaces for recognition,'' \emph{Journal of Cognitive Neuroscience}, vol. 3, no. 1, pp. 71----86, 1991.

\bibitem{huang}
T. S. Huang, \emph{PCA and its applications in image compression}, University of Illinois Lecture Notes, 1998.

\bibitem{horn}
R. A. Horn and C. R. Johnson, \emph{Matrix Analysis}, 2nd ed. Cambridge University Press, 2012.

\bibitem{gersho}
A. Gersho and R. M. Gray, \emph{Vector Quantization and Signal Compression}. Springer, 1992.

\end{thebibliography}

\end{document}